\documentclass[9pt,conference]{IEEEtran}
\IEEEoverridecommandlockouts
\ifCLASSINFOpdf
\usepackage{amsthm,amsmath}
\usepackage {xcolor}
\usepackage{graphicx}
\usepackage{xcolor}
\usepackage{amsfonts}
\usepackage{amssymb,mathtools}
\usepackage{algpseudocode}
\usepackage[linesnumbered,ruled,vlined]{algorithm2e}
\usepackage{nomencl}
\usepackage{tikz}

\newcommand{\circleNumber}[1]{%
  \begin{tikzpicture}[baseline=(char.base)]
    \node[shape=circle, fill=black, inner sep=2pt] (char) {\textcolor{white}{#1}};
  \end{tikzpicture}
}
\usepackage{cite}
\usepackage{textcomp}
\usepackage{hyperref}
\newtheorem{theorem}{Theorem}
\newtheorem{corollary}{Corollary}

\renewcommand{\IEEEbibitemsep}{0pt plus 0.5pt}
\makeatletter

\IEEEtriggercmd{\reset@font\normalfont\fontsize{7.9pt}{8.4pt}\selectfont}
\makeatother
\IEEEtriggeratref{1}

\def\BibTeX{{\rm B\kern-.05em{\sc i\kern-.025em b}\kern-.08em
    T\kern-.1667em\lower.7ex\hbox{E}\kern-.125emX}}
    
\begin{document}

\title{Risk-Aware and Explainable Framework for Ensuring Guaranteed Coverage in Evolving Hardware Trojan Detection
}




\author{
\IEEEauthorblockN{Rahul Vishwakarma}
\IEEEauthorblockA{\textit{Computer Engineering \& Computer Science Department} \\
\textit{California State University Long Beach}\\
Long Beach, CA, USA \\
rahuldeo.vishwakarma01@student.csulb.edu} 
\and
\IEEEauthorblockN{Amin Rezaei}
\IEEEauthorblockA{\textit{Computer Engineering \& Computer Science Department} \\
\textit{California State University Long Beach}\\
Long Beach, CA, USA \\
amin.rezaei@csulb.edu} 
}

\maketitle

\begin{abstract}
As the semiconductor industry has shifted to a fabless paradigm, the risk of hardware Trojans being inserted at various stages of production has also increased. Recently, there has been a growing trend toward the use of machine learning solutions to detect hardware Trojans more effectively, with a focus on the accuracy of the model as an evaluation metric. However, in a high-risk and sensitive domain, we cannot accept even a small misclassification. Additionally, it is unrealistic to expect an ideal model, especially when Trojans evolve over time. Therefore, we need metrics to assess the trustworthiness of detected Trojans and a mechanism to simulate unseen ones. In this paper, we generate evolving hardware Trojans using our proposed novel conformalized generative adversarial networks and offer an efficient approach to detecting them based on a non-invasive algorithm-agnostic statistical inference framework that leverages the Mondrian conformal predictor. The method acts like a wrapper over any of the machine learning models and produces set predictions along with uncertainty quantification for each new detected Trojan for more robust decision-making. In the case of a \textit{NULL} set, a novel method to reject the decision by providing a calibrated explainability is discussed. The proposed approach has been validated on both synthetic and real chip-level benchmarks and proven to pave the way for researchers looking to find informed machine learning solutions to hardware security problems.
\end{abstract}


\section{Introduction}
\label{Intro}
Hardware Trojan (HT) insertion is a malicious modification made to the design of a hardware component that can cause a device to malfunction, leak sensitive information, or even cause physical damage \cite{francq2015introduction}. As the semiconductor industry has been adopted a fabless model, the possibility of HTs being inserted at different manufacturing stages increases, presenting a significant security threat to hardware systems. Traditional HT detection techniques such as signature-based methods \cite{gbade2014signature} that analyze the Integrated Circuit (IC) functionality, layout, and timing, are often ineffective against sophisticated HT insertion attacks, especially when the Trojans can be designed to evolve over time. Therefore, there has been a growing shift towards the use of Machine Learning (ML)-based solutions for a more effective and efficient approach to detecting HTs. However, even after adhering to the ML evaluation best practices, it may backfire in the context of hardware security \cite{285613}. The majority of the existing ML-based solutions do not provide enough information on the dataset, where the distribution between classes differs substantially, and the evaluation of attacks under the concept drift or evolution of new incoming dataset. In this context, a detailed study has been conducted in \cite{quiring2022and} to understand if ML is the silver bullet for each and every problem in hardware security domain \cite{liu2021two}.

When using any ML method, one problem that we foresee is that even if a model claims to have a very small false positive rate, there is no guarantee that the same will be applicable for unseen data. This is due to concept drift, which is caused by the attacker's intelligently modified version of HT insertion techniques. One missed false positive can have not only a significant financial and economic impact but can also be life-threatening in high-risk, sensitive domains, such as implantable devices. Therefore, there is a need for additional metrics that can complement the existing model evaluation techniques, provide reliable decisions, and guarantee coverage for the predictions. In this work, we propose \textbf{PALETTE}, an ex\textbf{P}lainable fr\textbf{A}mework for evo\textbf{L}ving hardwar\textbf{E} \textbf{T}rojan de\textbf{TE}ction in risk-sensitive domains based on the algorithm-agnostic statistical inference technique of conformal prediction \cite{shafer2008tutorial} and provide risk-aware theoretical guaranteed coverage of predictions valid under covariate shift \cite{tibshirani2019conformal}. Our method is non-invasive which can be applied as a wrapper over existing ML models. In addition, rather than simply providing a point prediction, i.e., the detected circuit is infected with Trojan or not, it provides a set prediction of the detected labels resulting in the correct class being included 95\% of the time on average (i.e., $\alpha$ = 0.05).

\begin{figure*}[ht]
  \centering
   \includegraphics[width=0.95\linewidth]{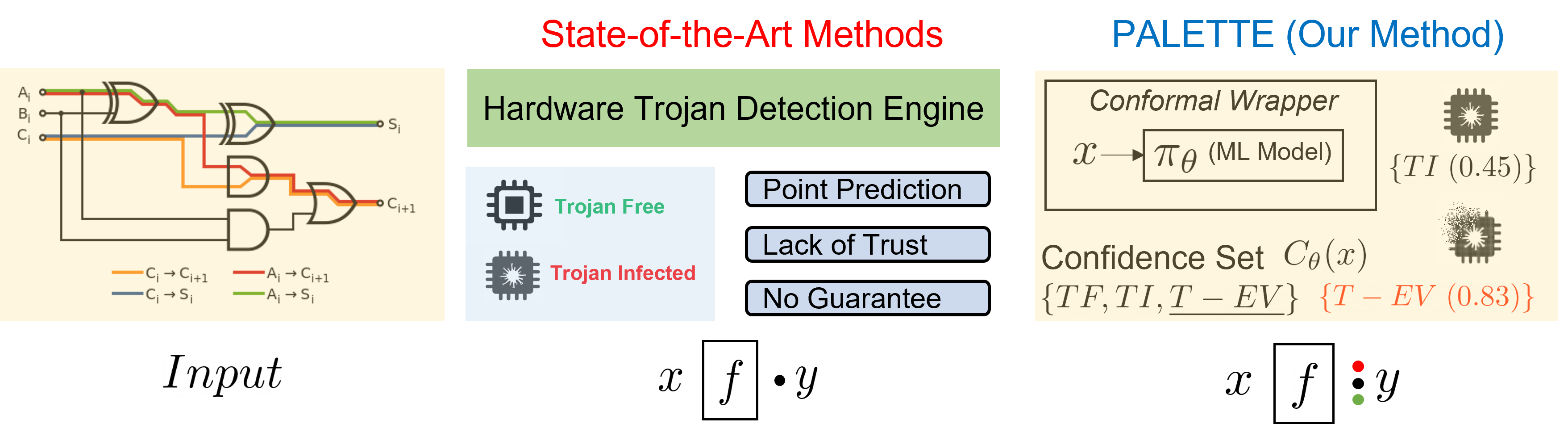}
   \caption{The input feature is passed to a conventional machine learning hardware Trojan detector. This detector produces a result and is compared to conformal inference, which provides an uncertainty measure and a confidence set for each detected evolved hardware Trojan.}
  \label{fig:intro}
\end{figure*}

In this paper, we pave the way for hardware security researchers to apply ML to their problems and create awareness about risk-controlled predictions with guaranteed coverage.

\subsection{Related Works}
\label{Related}
HT detection using traditional ML techniques has primarily focused on modeling methods, and it involves developing and implementing algorithms that can improve overall accuracy in detecting HTs. The input to the model consists of the features extracted from the Register Transfer Level (RTL) code as tabular and graphical representations of the circuit. Different surveys on ML for detecting HT attacks have been conducted in \cite{huang2020survey}, \cite{gubbi2023hardware}, \cite{koylu2023survey}, and \cite{kundu2021application}. In \cite{botero2021hardware} and \cite{ashok2022hardware} image classification is used, and in \cite{10027082} multimodal image processing is utilized. Most of the papers have focused on feature extraction from gate-level netlists and using ML models such as Support Vector Machine (SVM) \cite{bao2014application}, Neural Network (NN) \cite{hasegawa2017hardware}, eXtreme Gradient Boosting (XGB) \cite{dong2019machine}, and Random Forest (RF) classifier \cite{hasegawa2017trojan}.

With the success of Reinforcement Learning (RL) in other domains, a few works have adapted RL in the hardware security domain, such as RL-based static detection \cite{gohil2022attrition}, and RL with adaptive sampling for on-chip detection \cite{chen2023adatest}. In these methods, a common approach involves training the classifier model initially and then adjusting the hyperparameters of the classifier. This adjustment process aims to minimize the false negative rate and, in turn, improve the model's accuracy. Graph Neural Network (GNN) \cite{alrahis2022embracing,hepp2022golden} and Abstract Syntax Tree (AST) \cite{han2019hardware} are generated for the RTL code; however, it is still not clear how the graphs can carry forward not only the structural but also the behavioral attributes of the circuit.

Furthermore, dealing with concept drift is crucial after deploying a model because the newly acquired data can be significantly dissimilar to the data the model was originally trained on. One such work in security applications is shown in \cite{yang2021cade} which involves the mapping of data samples into a space with fewer dimensions and the automatic learning of a distance function that can evaluate the differences between them. The concept drift, however, has not been studied in the HT domain even though HTs can evolve over time.

The explainability aspect of ML is also covered in the hardware security domain; for example, \cite{pan2023hardware}, \cite{downing2021deepreflect}, and \cite{severi2021explanation} have used SHapley Additive exPlanations (SHAP) on benchmark datasets and have shown promising results. However, a drawback of utilizing SHAP is that it comes with inherent issues such as disregarding causality and being affected by human bias. It merely assesses the extent to which features contribute to a given dataset, without providing an explanation of how they would behave in the actual world, which may differ from the dataset used.

\subsection{Contributions}
\label{Contribution}
In this paper, we answer two questions. First, \textit{can we quantify the uncertainty associated with the HT prediction outcome and guarantee the validity of the predicted label with a handful of highly imbalanced data points? } Specifically, we are interested in the possibility that a ML classifier can predict the truth label of a new data point with 95\% provable guaranteed coverage, which is required in a risk-sensitive domain. Designing such a system can be beneficial to the decision maker who is going to investigate if the detected label is ``Trojan-Infected''. Second, \textit{can we rank the detected ``Trojan-Infected'' circuit for more informed treatment?} Answering this question will help the designer prioritize which one to take action on first for mitigation.

The big picture of our method compared with state-of-the-art ones is shown in Fig. \ref{fig:intro}. In this diagram, we also show the current limitations of using any of the ML frameworks for HT detection. First, they output the prediction as either label-A or label-B; second, there is a lack of trust in the prediction because not all ML classifiers are well calibrated; and third, there is no guarantee of coverage of the predictions. Our main contributions are as follows:

\begin{itemize}
\item Suggesting the notion of \textit{HT evolution} and providing an innovative method of creating evolving HTs with high precision by conformalized generative adversarial network.
\item Discussing the novel notion of \textit{guaranteed coverage} of the prediction set and proposing tunable significance level by leveraging the statistical inference tool of conformal prediction for HT detection.
\item Defining the notion of algorithm-agnostic and explainability-aware \textit{reject prediction} made by the ML model. When the model is uncertain of identifying the evolving Trojan, it simply says, ``I don't know,'' rejecting the prediction and passing it to a human for manual investigation.
\item Proposing a \textit{ranking mechanism} for the evolved Trojans by assigning the confidence score from the prediction and validating the proposed HT detection method both on synthetic and real chip-level HT-induced benchmarks.
\end{itemize}


\section{Preliminaries}
\label{Sec:Prelem}

\subsection{Calibrated Prediction}
\label{sec:Calibration}
Calibration involves ensuring that a model's confidence score accurately reflects the true probability of the prediction's correctness \cite{ovadia2019can}. Let $X$ be the input data, and $Y$ be the output label. Given a training dataset $D = {(x_1, y_1), (x_2, y_2),..., (x_n, y_n)}$, the goal is to learn a function $f$ that can predict the correct output label $y$ for a given input $x$. The output of the model for an input $x$ can be denoted as $f(x)$, and the true probability of the prediction's correctness can be denoted as $P(y=1|x)$. A calibrated model produces a confidence score $g(x)$ that reflects the true probability of correctness of the prediction. The goal of calibration is to ensure that the confidence score $g(x)$ is well-calibrated, i.e., $P(y=1|g(x)=p) = p$ for all $p$ in the range $[0, 1]$.

Calibration is a crucial aspect in HT detection since it aids in determining the likelihood of the existence of a Trojan in a circuit, which can have a significant impact on decision-making. In situations where a model's confidence score is high, but the likelihood of a Trojan's presence is low, it is reasonable to assume that the circuit does not contain a Trojan. Conversely, if the confidence score is low but the likelihood of a Trojan's presence is high, further investigation of the circuit is necessary.

\subsection{Conformal Prediction}
\label{sec:CP}

Conformal prediction \cite{shafer2008tutorial} is a ML framework that quantifies prediction uncertainty by generating prediction sets. It enhances the inference of traditional models, ensuring reliable validity and enabling confidence estimation for individual predictions. In the context of detecting HTs, label-conditional validity is a vital property when dealing with an imbalanced dataset where label proportions differ significantly. This is particularly relevant since the likelihood of encountering a Trojan on a circuit is generally low. In addition, it is worth noting that minority classes are often disproportionately impacted by errors when label-conditional validity is absent \cite{lofstrom2015bias}. However, this issue can be mitigated by ensuring label-conditional validity, which guarantees that the error rate for even the minority class will eventually converge to the chosen significance level in the long term. Sometimes, conformal prediction may produce uncertain predictions, meaning that prediction sets contain more than one value. This occurs when none of the labels can be rejected at the specified significance level.

When using conformal prediction, the confusion matrix differs slightly from the conventional one due to the unique nature of \textit{prediction sets}, which consist of multiple values rather than a single value. In the case of binary classification, it is essential to consider the number of correctly predicted examples, which have a prediction set containing only the correct label, as well as the number of incorrectly predicted examples, where the prediction set includes only the incorrect label. Additionally, it is important to take into account the number of inconclusive predictions that occur when the prediction set contains both labels, as well as the number of examples with an \textit{empty prediction set}. Furthermore, in some cases, it may be more appropriate to provide a \textit{single value point prediction} instead of a prediction set or interval in a hedged forecast. In such cases, selecting the label with the highest $p$-value is a simple and reasonable option. The point prediction can be hedged by incorporating additional information that describes the uncertainty.

Our work relies on Mondrian Inductive Conformal Prediction (ICP) \cite{bostrom2021mondrian} in Algorithm \ref{algo:mcp} and to decrease the rate of false negatives in alert systems, we require class-based authenticity for samples classified as ``Evolving Trojan''. Additionally, we must ensure that the samples labeled as``Evolving Trojan'' are indeed genuine to attain this goal.

\begin{algorithm}[t]
\SetKwInOut{Input}{Input}
\SetKwInOut{Output}{Output}
\Input{Training data $D$, test instance $x$, significance level $\alpha$, number of trees $T$, and maximum tree depth $d$.}
\Output{Prediction set $C(x)$ for $x$.}
Divide $D$ into $T$ disjoint subsets $D_1, \ldots, D_T$;

\For{$t \leftarrow 1$ \KwTo $T$}{
Sample $D_t'$ from $D_t$ by recursively partitioning $D_t$ along randomly chosen hyperplanes until each partition contains at most $2d$ points.

Train a classification model $M_t$ on $D_t'$.
}
Compute the conformity scores $s_t(x)$ of $x$ with respect to each model $M_t$.

Sort the conformity scores $s_t(x)$ in decreasing order.

Compute the $p$-values $p_t$ of the $T$ conformity scores $s_t(x)$ using the formula $p_t = \frac{T - t + 1}{T}$.

Compute the threshold $h$ such that $h = s_t(x)$ if $p_t > \alpha$, otherwise $h = \infty$.

Construct the prediction set $C(x)$ as the set of all labels $y$ such that $s_t(y) \geq h$ for all models $M_t$.

\Return $C(x)$
\caption{Mondrian ICP}
\label{algo:mcp}
\end{algorithm}

When calculating the non-conformity scores, we only consider the scores related to the examples that share the same class as the object $x_{n+1}$, which we are testing hypothetically as shown below: 
$$
p_{n+1}^{C_k}=\frac{\left|\left\{i \in 1, \ldots, q: y_i=C_k, \alpha_{n+1}^{C_k} \leq \alpha_i\right\}\right|}{\left|\left\{i \in 1, \ldots, q: y_i=C_k\right\}\right|}
$$

\subsection{Guaranteed Coverage of Prediction}
\label{sec:Coverage}
In the domain of HT detection, it is not only important to have a high level of confidence in the predictions made by a model but also a guarantee of the coverage of each prediction. The property of guaranteed coverage is an inherent property of conformal prediction, which provides statistical guarantees of the correctness of the model's predictions \cite{angelopoulos2023conformal}. The theoretical guarantee of coverage is based on the significance level, which is the probability of the model making a mistake. For example, if we set the significance level to $0.05$, it means that we allow the model to make mistakes $5\%$ of the time.

The theoretical guarantee of coverage is valid for any input $x$, that the true output label $y$ will be contained in the prediction set $C(x)$ with a probability of at least $1 - \alpha$, where $\alpha$ is the significance level. Mathematically, this can be expressed as:

$$
P(y \in C(x)) \geq 1 - \alpha
$$

In other words, the probability of making a mistake is bounded by $\alpha$, and as $\alpha$ decreases, the size of the prediction set decreases, leading to higher confidence in the model's predictions. For example, if we set $\alpha = 0.05$, it means that we are $95\%$ confident that the true output label $y$ is contained in the prediction set $C(x)$ for any input $x$. The use of conformal prediction provides a strong theoretical guarantee of the correctness of the model's predictions in the context of HT detection, and the corresponding proof is given in Theorem \ref{thm:coverage}.

\begin{theorem}\label{thm:coverage}
Let $\mathcal{D}$ be a probability distribution over a set $\mathcal{X} \times \{0,1\}$, where $\mathcal{X}$ is a set of input features and $\{0,1\}$ is the set of labels. Let $f: \mathcal{X} \to \{0,1\}$ be a binary classifier, and let $\epsilon \in (0,1)$ be a confidence level. Then, the conformal prediction algorithm outputs a set of predictions $C(x) \subseteq \{0,1\}$ for each input $x \in \mathcal{X}$ such that:

\begin{equation*}
\mathbb{P}[(x,y) \sim \mathcal{D}, y \in C(x)] \geq 1-\epsilon
\end{equation*}

where $(x,y) \sim \mathcal{D}$ denotes sampling a pair $(x,y)$ from the distribution $\mathcal{D}$.

\end{theorem}

\begin{proof}
The proof follows from the construction of the conformal prediction algorithm. Given an input $x$, the algorithm outputs a set of predictions $C(x)$ based on the observed labels of the training examples with similar input features to $x$. The algorithm guarantees that each prediction in $C(x)$ has a $p$-value less than or equal to $\epsilon$ for any new input with the same feature vector as $x$. Since the algorithm outputs a set of predictions, the probability that at least one of the predictions is correct is at least $1-\epsilon$. 
\end{proof}

\begin{corollary}\label{cor:coverage}
Let $\mathcal{D}$, $f$, and $\epsilon$ be as in Theorem~\ref{thm:coverage}. For any sample size $n$, the conformal prediction algorithm outputs a set of predictions $C(x_1),\ldots,C(x_n)$ for each input $x_1,\ldots,x_n \in \mathcal{X}$ such that:

\begin{equation*}
\mathbb{P}[\forall i \in \{1,\ldots,n\}, (x_i,y_i) \sim \mathcal{D}, y_i \in C(x_i)] \geq 1-\epsilon
\end{equation*}

where $(x_i,y_i) \sim \mathcal{D}$ denotes sampling a pair $(x_i,y_i)$ from the distribution $\mathcal{D}$ for each $i$.

\end{corollary}

\begin{proof}
The proof follows from a union bound over the $n$ samples:

\begin{equation*}
\begin{aligned}
&\mathbb{P}[\forall i \in \{1,\ldots,n\}, (x_i,y_i) \sim \mathcal{D}, y_i \in C(x_i)] \\
&\geq 1 - \sum_{i=1}^n \mathbb{P}[(x_i, y_i) \sim \mathcal{D}, y_i \notin C(x_i)] \\
&\geq 1 - n \epsilon
\end{aligned}
\end{equation*}

where the second inequality follows from Theorem~\ref{thm:coverage}. 
\end{proof}

\section{Notion of Evolution \& Hardware Trojans} 
\label{sec:Evolution}
Darwin, in his book, \textit{On the Origin of Species}, referred to ``descent with modification'', instead of \textit{evolution}. Further, a more expansive definition of evolution was given by Futuyma \cite{laland2014does}: ``biological evolution is change in the properties of groups of organisms over the course of generations; it embraces everything from slight changes in the proportions of different forms of a gene within a population to the alterations that led from the earliest organism to dinosaurs, bees, oaks, and humans''. Now, we narrow down the notion of evolution for HTs based on the following assumptions:

\begin{itemize}
    \item The structural (genotype) and behavioral (phenotype) characteristics of HTs change over a period of time, and the changes are induced by the attacker;
    \item Structural changes can be mathematically formulated for the evolved Trojan as \[ E_{HT} \rightarrow HT \blacksquare HT_{structural\_changes} \] 
    where HT is an existing Trojan and $\blacksquare$ is the operation for structural changes which creates an evolved Trojan $E_{HT}$;
    \item Behavioral changes are mapped with natural selection, which is the driving force for evolution. The attacker designs the HT such that it adapts to the IC (ecosystem) and its malicious impact is not easily detectable on the circuit. (i.e., it increases its chance of survival.)
\end{itemize}

We use the above assumption to include the notion of evolution and derive an evolved dataset in a ML-based HT detection engine. In the context of HTs, we can either detect the evolution or predict the evolution way ahead of time. The detection can be performed using anomaly detection \cite{liu2022anomaly}; however, here we will be focusing on the prediction of evolution. If we can predict the evolutionary changes in the dataset, a specific treatment can be performed to mitigate the impact of HT insertion. To the best of our knowledge, we have not come across any work in the literature that considers the evolutionary aspect while designing HT detection approaches.

The evolutionary dataset optimization discussed in \cite{edo-paper} optimizes any real-valued function over a subset of the space of all possible datasets. It is not feasible to adapt this method for our use case as our real-time data will be Non-Independent and Identically Distributed (Non-IID). An alternate approach can be to use evolutionary algorithms, as discussed in Box2d \cite{catto2010box2d}. There, the problem statement is to evolve the structure of a toy car, provided the geometry of the car shape is translated to chromosomes. The issue with this approach is that we should know how the evolved car looks; however, in our case, we never know the structure of the evolved Trojan.

\begin{algorithm} [t]
\SetKwInOut{Input}{Input}
\SetKwInOut{Output}{Output}
\Input{Training dataset $\mathcal{D} = \{(x_i, y_i)\}_{i=1}^n$, where $x_i \in \mathbb{R}^p$ and $y_i \in \{0, 1\}$ are the feature vector and label for the $i$-th example, respectively; significance level $\alpha$; number of conformal predictors $M$; GAN generator $G$; discriminator model $D$}
\Output{Conformalized discriminator model $D_\text{CP}$}

\For{$m=1$ \KwTo $M$}{
    $\mathcal{D}_m \gets$ bootstrap sample of $\mathcal{D}$\;
    Train GAN generator $G_m$ on $\mathcal{D}_m$\;
    Generate synthetic dataset $\mathcal{D}_\text{synth}^m = \{G_m(z_i)\}_{i=1}^n$, where $z_i \in \mathbb{R}^k$ are random noise vectors\;
    Train discriminator model $D_m$ on $\mathcal{D}_m \cup \mathcal{D}_\text{synth}^m$\;
}

\For{$i=1$ \KwTo $n$}{
    $X_i \gets \{x_i\} \cup \{G_m(z_i)\}_{m=1}^M$, where $z_i \in \mathbb{R}^k$ are random noise vectors\;
    $\text{CP}_i \gets$ conformal predictor trained on $(X_i, y_i)$ with significance level $\alpha$\;
    $p_i \gets \text{CP}_i(D(x_i))$\;
}

Train conformalized discriminator model $D_\text{CP}$ on $\{(x_i, y_i, p_i)\}_{i=1}^n$\;

For each sample $x_i$ in the test set $D_{test}$, make a prediction based on whether $D(x_i)$ is within the prediction interval $I_i$:
\begin{equation*}
y_i =
\begin{cases}
1 & \text{if } D(x_i) \notin I_i \\
0 & \text{if } D(x_i) \in I_i
\end{cases}
\end{equation*}

\Return{$D_\text{CP}$}

\caption{Conformalized GAN}
\label{alg:cgan1}
\end{algorithm}

\subsection{Genetic Algorithm} 
\label{sec:GA}
Genetic Algorithms (GAs) \cite{holland1992genetic} have been used to evolve the architecture of NNs for understanding the security of logic locking \cite{sisejkovic2021challenging}. The most challenging part of using GA is designing a fitness function. In our case, one possible design of fitness can focus on the ensemble efficiency of detection methods and then compare the similarity of the child Trojan with the list of HTs in a dictionary. However, the limitation of this fitness function is that it will never be able to estimate the fitness of Trojans that are out of distribution.

\begin{figure*}[ht]
  \centering
   \includegraphics[width=1\linewidth]{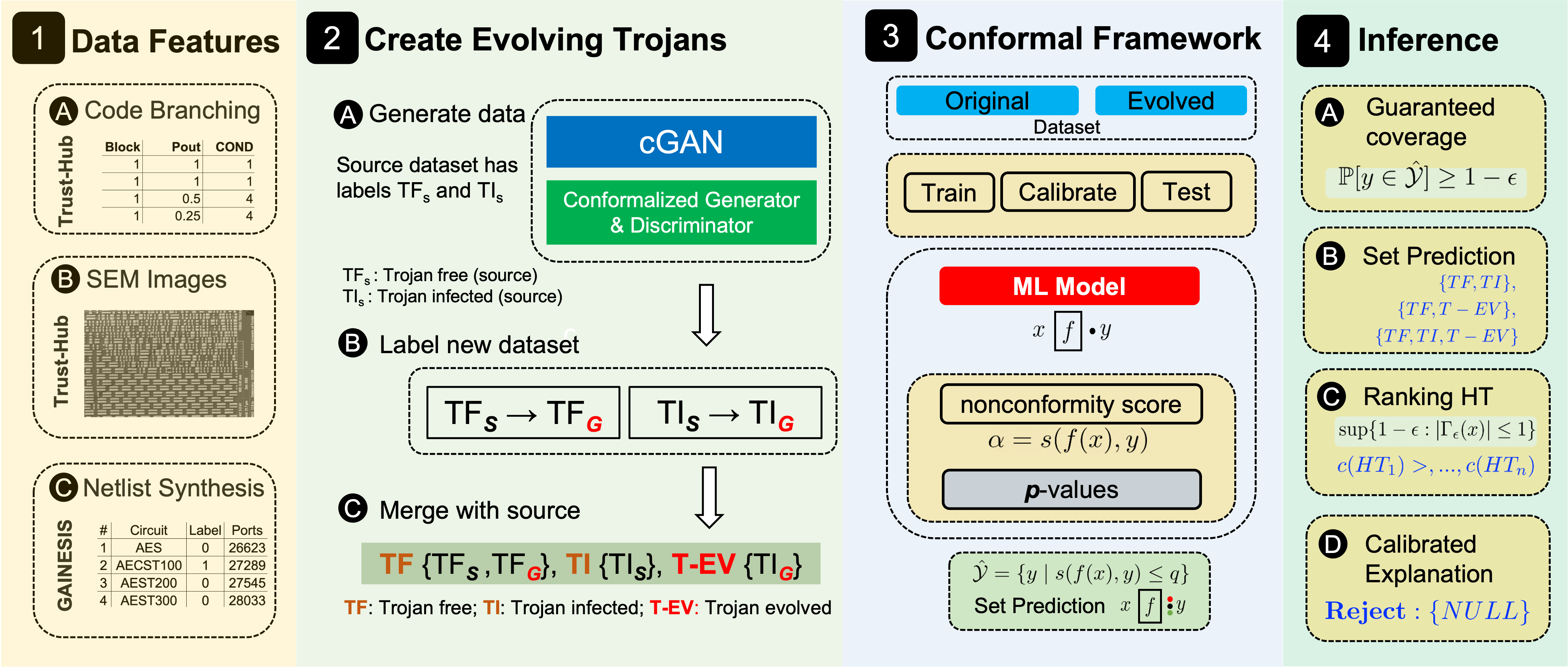}
  \caption{PALETTE: Proposed solution showing the method to design evolving hardware Trojan by tuning conformalized generative adversarial network and using the evolved dataset to make informed, risk-aware predictions with guaranteed coverage.}
  \label{fig:solution}
\end{figure*}

\subsection{Generative Adversarial Network} 
\label{sec:GAN}
Based on the game theory and optimization approach, the objective of generative modeling \cite{goodfellow2020generative} is to analyze a set of training examples and acquire knowledge about the likelihood distribution that created them. Generative Adversarial Network (GAN) has been successfully used for detecting fake images \cite{ojha2023towards} and text-to-image synthesis \cite{kang2023scaling}. In the recent past, there has been a shift in focus towards the utilization of GANs for working with tabular data. An instance of this approach is used for conditional GAN, as demonstrated by \cite{xu2019modeling}, which models tabular data and is also effective with imbalanced data. 

Here are three prime motivations for using GAN for synthesizing the HT dataset. 

\subsubsection{Highly Imbalanced Data} In a real-time scenario, the labels for Trojan-Infected circuits are very rare and difficult to detect. This gives rise to the problem of an imbalanced dataset. Based on the existing literature, we believe GANs can be used to generate a more realistic synthetic dataset that complements the training phase.

\subsubsection{Non-IID Case for Law of Large Number} The evolved Trojan may or may not be from the same distribution, and for this reason we have to consider the case of Non-IID random variables. One such an example is demonstrated in \cite{yonetani2019decentralized}. We also know that for a large enough dataset with Non-IID samples, the sample mean will converge to the true population mean as the sample size increases. This implies that as the size of the dataset increases, the statistical properties of the data become more reliable and consistent. Thus, the larger the dataset, the more accurate the model is likely to be, provided that there are enough computational resources to effectively process the data. The proof given in \cite{etemadi1981elementary} for the strong law of large numbers can be generalized to an r-dimensional array of random variables where the sufficient condition becomes $E\left(|X|\left(\log ^{+}|X|\right)^{r-1}\right)<\infty$ based on the theorem and corresponding proof for Non-IID given in \cite{smythe1973strong}. The Non-IID case is worthy of our attention, as evolved HT might not represent the same distribution of population in real-time.

\subsubsection{Risk Sensitive Application} Given the potential for significant financial losses, we cannot afford to tolerate even a small probability of a false positive. To mitigate this, we start by designing a near-realistic synthetic dataset using GAN by conformalizing the discriminator and generator.

\section{Designing \& Predicting Evolving Hardware Trojans}
\label{sec:Solution}
Our proposed evolving HT detection method called \textbf{PALETTE} is shown in Fig. \ref{fig:solution} with four major components.

\circleNumber{1} As with any ML-based solution, the first step is to extract the dataset, and in the case of HTs, we can have images, tables, and graphs as input dataset for HT classification. For example, the features extracted from an IC can be Scanning Electron Microscope (SEM) images, as used in \cite{vashistha2018trojan,shi2019golden}. 
In our case, we have used the features extracted based on code branching from the TrustHub chip-level Trojan dataset \cite{px6s-sm21-22} and the netlist synthetic dataset based on GAINESIS \cite{liakos2022gainesis}. 

\begin{figure}[!b]
  \includegraphics[width=1\linewidth]{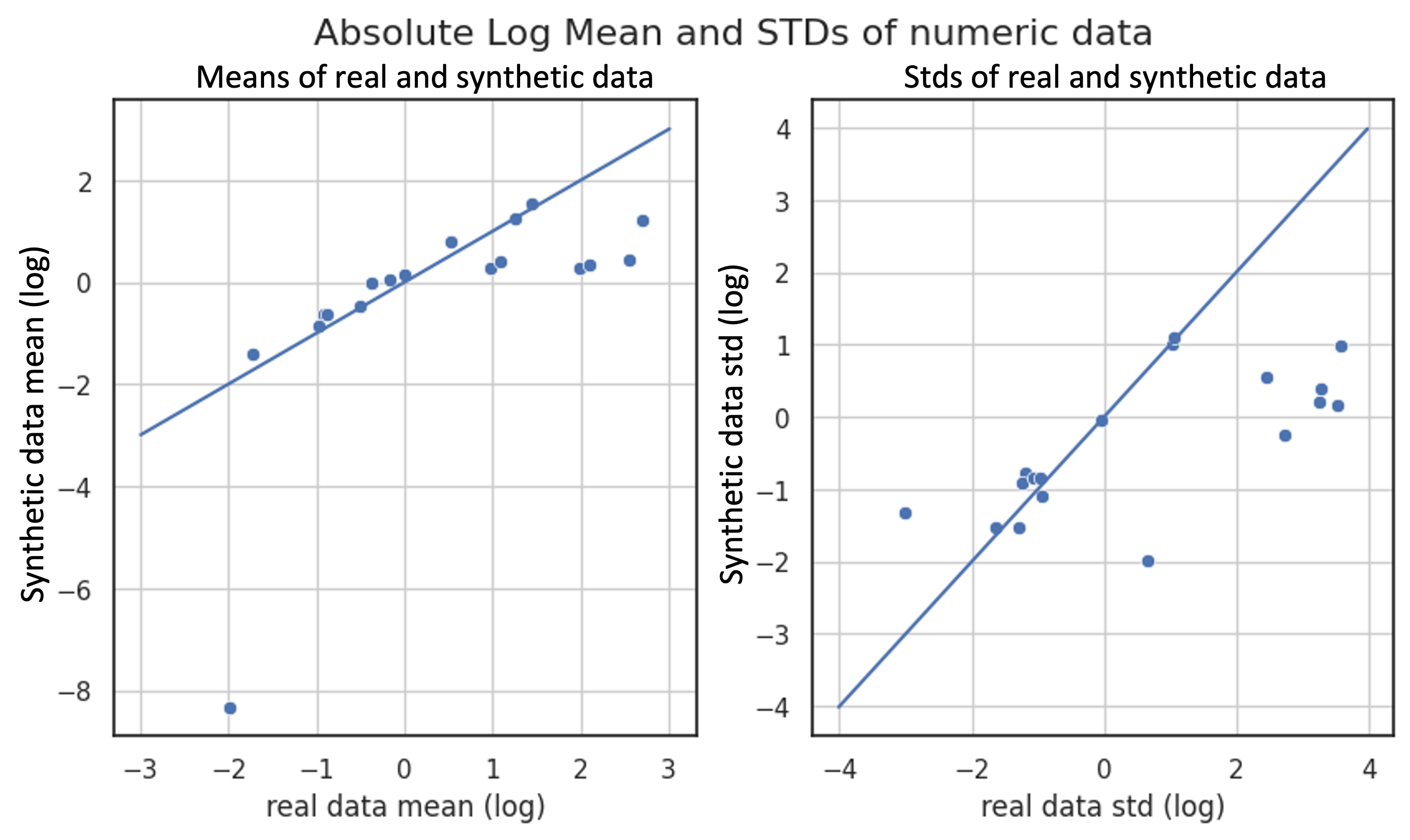}
  \caption{Comparison of real and synthetically generated dataset on Trust-Hub chip-level Trojan dataset.}
  \label{fig:gan1}
\end{figure}

\circleNumber{2} We introduce the Conformalized GAN algorithm, which is illustrated in Algorithm \ref{alg:cgan1}. Our algorithm is inspired by \cite{sankaranarayanan2022semantic}, which leverages principled uncertainty intervals to generate high-quality images from corrupted inputs, and the uncertainty intervals provide a guarantee of containing the true semantic factors for any underlying generative model. Motivated by their work, we generate high-quality evolved representation of HTs from existing Trust-Hub dataset \cite{px6s-sm21-22} and the netlist synthesis GAINESIS dataset \cite{liakos2022gainesis}. 
The algorithm utilizes conformal prediction to generate evolving HTs and determine its associated level of confidence using prediction intervals. A comparison of Trust-Hub source dataset with the synthetically generated data, which we call the evolved dataset, is shown in Fig. \ref{fig:gan1}. In contrast with traditional GANs our proposed method provides a more reliable means for generating evolving HTs.  

\circleNumber{3} The dataset is further fed as input to the conformal inference engine, which outputs set prediction instead of point predictions based on the significance level. The method is \textit{algorithm-agnostic} as any ML classifier such as statistical or deep learning can be used as shown in Fig. \ref{fig:solution}. The non-conformity score is calculated for each prediction. The $p$-value represents the probability that the prediction is correct and is used to determine the guaranteed coverage. The important part of the solution is how we interpret the results in a risk sensitive domain where we cannot tolerate even a single wrong decision.

\circleNumber{4} We derive four different inferential use cases based on conformal inference. The motive is to quantify the uncertainty associated with each prediction and reduce the False Discovery Rate (FDR) for Trojan-Free (TF), Trojan-Infected (TI), or Evolving Trojan (T-EV). The first is guaranteed coverage, which claims that based on the user-defined significance level, the predicted label will belong to that class. Here, considering the degree of risk associated with the prediction, a significance level is defined and applied to the $p$-values of each label for the data point of the circuit. The second is an inherent property of conformal prediction that results in a set prediction which can have all the labels \{TF, TI, T-EV\}, a combination of labels \{TF, T-EV\} or \{TI, T-EV\}, or a single label \{TF\}, \{TI\}, or \{T-EV\}. The third, ranks the predicted HTs by calculating the confidence of each prediction and using it to rank the severity of being infected with a Trojan (TI, T-EV). The purpose of ranking is to prioritize which one to take action on first for mitigation. Finally, the fourth is calibrated explanation for the predictions where the model says: \textit{``I don't know''} and rejects the prediction. The proposed method overcomes the issues of local explanations by SHAP as discussed in Section \ref{Related} and provides a calibrated approach to reasoning out \textit{``why''} a certain prediction has to be rejected. This is achieved by a \textit{NULL} set, indicating that the model is not able to output the prediction for a specific significance level (1 - $\alpha$). These four risk-aware, tailored prediction use cases are discussed with experimental results in Section \ref{sec:Results}.

\section{Experimental results}
\label{sec:Results}
In this section we share the experimental results on two datasets. First is GAINESIS \cite{liakos2022gainesis} synthetic dataset with binary labels and second is using Trust-Hub chip-level Trojan dataset \cite{px6s-sm21-22}. This dataset includes VHDL or Verilog source code files for each IP core design, which contain both malicious and non-malicious functions. The malicious functions are often embedded within conditional statements that are seldom executed. Consequently, the ML features are extracted from these conditional statements. We used Python (3.9) and implemented the solution on macOS (13.3.1) having 8 GB RAM with built-in GPU. The experimental results with source code and the dataset are hosted on GitHub \footnote{https://github.com/cars-lab-repo/PALETTE/}.

\subsection{Evolved Dataset}
\label{Sec:EvolvedDataset}
We first generated 10,000 data points using the proposed conformalized GAN  with the given source dataset and picked only 20\% of the evolved dataset. The generated dataset has labels $TF_{G}$ and $TI_{G}$, where as the source dataset has labels $TF_{S}$ and $TI_{S}$. In our evolved dataset, we create three labels as shown below. First, Trojan-Free (TF) which consists of $TF_{S}$ and $TF_{G}$; second, Trojan-Infected (TI) where we only consider the label $TI_{S}$; finally, the third label is Evolved Trojan (T-EV) which consists of the label $TI_{G}$. 
$$Label = \{TF, TI, T-EV \}$$

The dataset is split into training set, calibration set, and test set with ratio 2:1:1. In training dataset we have 1436 TF, 114 TI, and 308 T-EV. For calibration, we have 470 TF, 33 TI, and 117 T-EV. Finally, we have 18\% of T-EV in calibration set and 16\% each in train and test.

The dataset is split into training set, calibration set, and test set as shown in Table \ref{tab:dataset}. 

\subsection{Baseline Model}
\label{Sec:Base}
We can choose any of the classification algorithms as a baseline model because \textbf{PALETTE}, as described in Section \ref{sec:CP}, is algorithm-agnostic. Here, we have used logistic regression as a classifier to detect the evolving HTs, and we evaluate the accuracy of the models as a performance metric. If we use logistic regression to detect HTs, the overall accuracy is 0.85, while if we use conformal inference as a wrapper over the logistic regression, the accuracy increases to 0.88 for $\alpha$ = 0.05 and 0.90 for $\alpha$ = 0.1. This also shows the performance improvement of any classification model when used with underlying conformal inference. A detailed result is shared on our GitHub repository.

\begin{figure}[b]
  \includegraphics[width=1\linewidth]{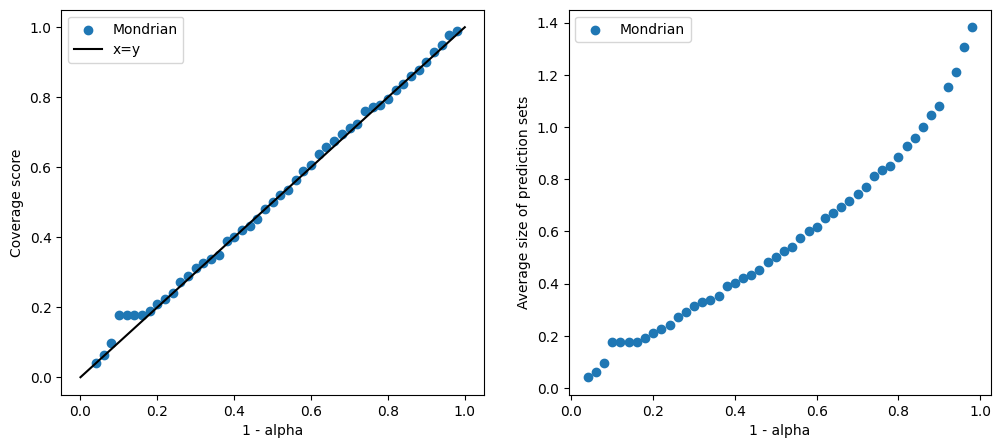}
  \caption{Effective coverage and average prediction set size for Trust-Hub chip-level Trojan dataset.}
  \label{fig:split}
\end{figure}

\subsection{Conformal Inference}
\label{Sec:ConformalInference} 
We will emphasize and reiterate that \textbf{`any'} classification algorithm can be used along with the conformal inference framework, and in our work we have adopted logistic regression with Mondrian conformal predictors. The $p$-value is a measure of confidence in the predictions made by the ML model. It is like a score that tells us how well the model is doing when it makes predictions about new data. To calculate the $p$-value, we compare the model's prediction for a new piece of data with its predictions for the data it was trained on based on hypothesis testing. If the new data is very different from what the model has seen before, the $p$-value will be small, and this can be a sign that the model's prediction for the new data might not be as accurate. So, we need to be careful when interpreting predictions from the model if the $p$-value is too small.

\begin{figure*}[]
\centering
  \includegraphics[width=\linewidth]{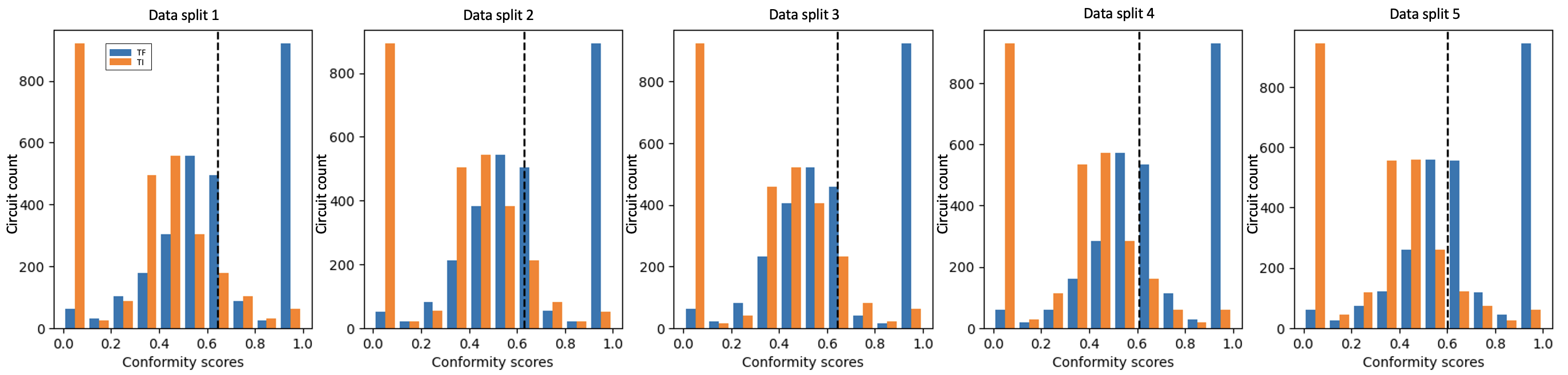}
  \caption{Distribution of scores on each five of the calibration fold for the Mondrian conformal predictor for GAINESIS dataset.}
  \label{fig:b_fold}
\end{figure*}

\begin{table}[t]
\centering
\caption{Conformal inference and associated p-vales for Trust-Hub chip-level Trojan dataset.}
\begin{tabular}{lcccrrrrr}
\hline
  & \textbf{TF} & \textbf{TI} & \textbf{T-EV} & \textbf{pTF} & \textbf{pTI} & \textbf{pT-EV} & \textbf{y\_pred} & \textbf{Conf}\\ \hline
1 & T           & F           & F           & 0.319        & 0            & 0.003        & TF  & 0.997             \\ \hline
2 & T           & F           & F           & 0.243        & 0.002        & 0.006        & TF  & 0.994             \\ \hline
3 & T           & T           & F           & 0.161        & 0.078        & 0.016        & TF   & 0.992            \\ \hline
4 & T           & T           & T           & 0.114        & 0.053        & 0.119        & T-EV  & 0.886           \\ \hline
5 & T           & F           & F           & 0.645        & 0.001        & 0.004            & TF   & 0.996           \\ \hline
6 & F           & F           & T           & 0.653            & 0            & 0.971        & T-EV    & 0.365         \\ \hline
7 & T           & F           & F           & 0.3          & 0            & 0.002        & TF      & 0.998         \\ \hline
\end{tabular}
\label{tab:conformalresults}
\end{table}

\begin{table}[t]
\centering
\caption{Conformal inference for GAINESIS dataset.}
\begin{tabular}{lllll}
\hline
\textbf{circuit}        & \textbf{TI}             & \textbf{TF}             & \textbf{y-pred} & \textbf{Conf}           \\ \hline
1                       & FALSE                   & TRUE                    & TF              & 0.891                   \\ \hline
2                       & FALSE                   & TRUE                    & TF              & 0.796                   \\ \hline
3                       & FALSE                   & TRUE                    & TF              & 0.996                   \\ \hline
4                       & FALSE                   & TRUE                    & TF              & 0.997                   \\ \hline
\multicolumn{1}{c}{...} & \multicolumn{1}{c}{...} & \multicolumn{1}{c}{...} & ...             & \multicolumn{1}{c}{...} \\ \hline
4596                    & FALSE                   & TRUE                    & TF              & 1                       \\ \hline
4597                    & FALSE                   & TRUE                    & TF              & 0.991                   \\ \hline
4598                    & TRUE                    & FALSE                   & TI              & 0.995                   \\ \hline
4599                    & FALSE                   & TRUE                    & TF              & 0.989                   \\ \hline
4600                    & FALSE                   & TRUE                    & TF              & 0.992                   \\ \hline
\end{tabular}
\label{tab:binary_cp}
\end{table}

The results obtained after implementing conformal inference for detecting evolving HTs are shown in Table \ref{tab:conformalresults}. Each row represents the circuit and the truth value in columns TF, TI, and T-EV. In addition, the $p$-values for each label are mentioned in the columns pTF, pTI, and pT-EV. Finally, the detected Trojan is mentioned in column y\_pred with $\alpha$ = 0.05. The column \textit{Conf} represents the confidence score of each detected label for each circuit, which is obtained by $1 - 2^{nd}p_{max}$. An application of conformal inference is the improvement of detection quality for evolving HTs. For example, in Table \ref{tab:conformalresults}, circuit 2 is detected as Trojan-Free because the $p$-values of TI and T-EV are less than the value of $\alpha$ = 0.05. The circuit 4 and 6 are detected as infected with an evolved Trojan. In circuit 4, we see that $p$-values for TF, TI, and T-EV are greater than the value of $\alpha$, so all the labels are set as T (True), and the maximum of the $p$-value is specified for the detected label. For example, with conformal inference, we can say that with 95\% detection guarantee (as $\alpha$ = 0.05 decided by the user), circuit 4 is detected as an evolving Trojan with a confidence score of 0.886. This helps the end user have granular-level reasoning for trustworthy and robust decision-making.

We also share the results for binary labels (TF, TI) on the GAINESIS dataset in Table \ref{tab:binary_cp}. The method was validated on $4600$ synthetic circuits with and without Trojans, and the corresponding confidence score is shown in the column \textit{Conf}.

\begin{table}[t]
\centering
\caption{Comparison of conformal predictors with corresponding significance level on Trust-Hub chip-level Trojan dataset.}
\begin{tabular}{ccccc}
\hline
\textbf{alpha} & \textbf{mondrian} & \textbf{raps} & \textbf{naïve} & \textbf{top\_k} \\ \hline
0.05           & 10                & 37            & 35             & 0               \\ \hline
0.5            & 45                & 57            & 57             & 61              \\ \hline
0.9            & 45                & 61            & 61             & 61              \\ \hline
\end{tabular}
\label{tab:racp}
\end{table}

Furthermore, we also explored variations of conformal predictors as described in \cite{bates2021distribution}. Table \ref{tab:racp} shows that the Mondrian conformal predictor is very strict on detecting the evolved hardware Trojans as compared to the risk-adaptive prediction set \textit{raps}, \textit{naive}, and \textit{top\_k} methods with varying significance levels. The \textit{naive} and \textit{top\_k} first get the model output of the true class, and \textit{naive} makes the estimated set prediction by getting quantiles from the score distribution, while \textit{top\_k} gets the quantiles from the distribution of the ordered positions of the true label. The \textit{raps} method first sorts the model output in decreasing order to get cumulative output of the true class and then uses it to obtain quantiles from cumulative score distributions. Furthermore, with a very high coverage of 95\% ($\alpha$ = 0.05), \textit{raps} and \textit{naive} detect almost three times more Trojans as compared to Mondrian, while the detection coverage becomes almost similar when the coverage level is increased.


\subsection{Performance Metrics}
\label{Sec:Performance}
Unlike classification task which produces Receiver Operating Characteristic (ROC) and Area Under Curve (AUC), conformal inference produces \textit{effective coverage} and \textit{efficiency}, i.e., average prediction set size, as performance metrics. The limitation of ROC and AUC is that they can be impacted by an imbalanced dataset. In Fig. \ref{fig:split} we show the two different performance metrics for Mondrian conformal predictors. The coverage score measures the proportion of instances in which the true label falls within the predicted region. It is typically measured at different confidence levels. Higher coverage indicates a more conservative prediction method. Now, since validity is guaranteed for all conformal predictors, the key performance metric is efficiency, i.e., the size of the label sets, where smaller sets are more informative and indicate higher efficiency. It is also a direct measure of how good the conformal predictor is at rejecting class labels.

When evaluating conformal prediction methods, there are several metrics that can be used to assess their performance. In Table \ref{tab:performance} we show the various performance metrics associated with the detection mechanism for significance levels ranging from 0.05 to 0.9. For example, avg\_c indicates the average number of class labels in the prediction sets; this metrics serves as a straightforward indicator of the conformal predictor's ability to accurately discard class labels.


The significance level is like a threshold that controls how often the ML model makes incorrect predictions. If we set a higher significance level, the model will make fewer errors, but its predictions may be less precise. So, we need to find the right balance to get the best results from our model.

Furthermore, we also show the performance metrics for the GAINESIS dataset in Fig. \ref{fig:b_fold}. We examine the conforming score (expected label) distribution on each of the five calibration folds for the Mondrian conformal predictor and observe that there is no major difference in the conforming score for each calibration split.

\begin{table}[t]
\centering
\caption{Performance metrics of conformal inference on Trust-Hub chip-level Trojan dataset.}
\begin{tabular}{lllll}
\hline
\textbf{sig} & \textbf{mean\_err} & \textbf{avg\_c} & \textbf{n\_correct} & \textbf{mean\_T-EV} \\ \hline
0.05         & 0.049                 & 1.040           & 589                 & 0.012               \\ \hline
0.1          & 0.102                 & 0.941           & 556               & 0.045               \\ \hline
0.2          & 0.204                 & 0.812           & 493             & 0.133               \\ \hline
0.3          & 0.303                 & 0.701           & 431             & 0.220               \\ \hline
0.4          & 0.406                 & 0.596           & 367            & 0.319               \\ \hline
0.5          & 0.504                 & 0.497           & 307           & 0.423               \\ \hline
0.6          & 0.604                 & 0.397           & 245             & 0.536               \\ \hline
0.7          & 0.702                 & 0.298           & 184            & 0.650               \\ \hline
0.8          & 0.798                 & 0.202           & 125           & 0.764               \\ \hline
0.9          & 0.900                 & 0.100           & 61         & 0.884               \\ \hline
\end{tabular}
\label{tab:performance}
\end{table}

\begin{figure*}[h]
\centering
  \includegraphics[width=\linewidth]{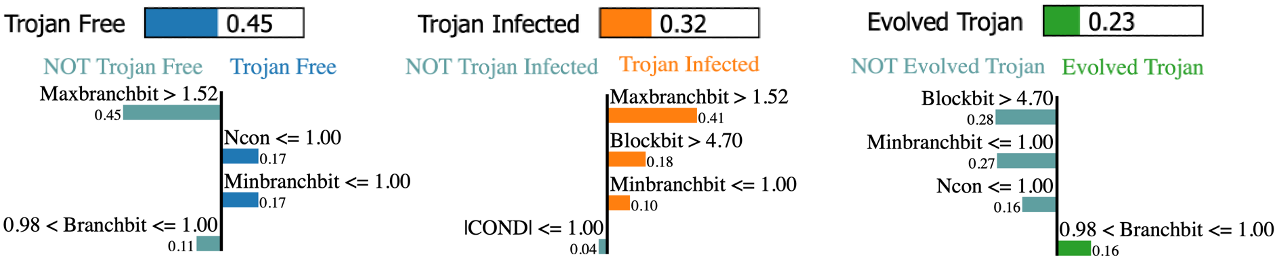}
  \caption{Calibrated explanation for rejecting a decision.}
  \label{fig:lime}
\end{figure*}

\subsection{Risk-Aware Ranking}
\label{Sec:Risk}
We leverage \textit{confidence score} from the conformal inference as a ranking mechanism for evolved HTs. For the below given circuits 12, 13, and 14 from the Trust-Hub dataset, we calculated their confidence score (C) with $\alpha$ = 0.05. 
$$
\begin{aligned}
& \alpha_{0.05}(\text { circuit } 12)=\{T-E V\}_{C=0.88} \\
& \alpha_{0.05}(\text { circuit } 13)=\{T-E V\}_{C=0.81} \\
& \alpha_{0.05}(\text { circuit } 14)=\{T-E V\}_{C=0.61}
\end{aligned}
$$
Confidence in a model's prediction is determined by its $p$-value which indicates the probability of obtaining a similar outcome under the \textit{NULL} hypothesis. Higher confidence implies greater accuracy. This metric is defined as:
$$\text{Confidence} (x) = \sup \{1 - \epsilon : |\Gamma_\epsilon (x)| \leq 1\} $$

By ranking the predictions, conformal prediction can offer a more informative way to assess the reliability of individual predictions. The ranked list allows decision-makers to set thresholds or confidence levels for accepting or rejecting predictions based on their position in the ranking.

This provides a flexible tool for controlling the trade-off between accuracy and reliability in different applications. In Table \ref{tab:confcred} we show the \textit{confidence} and \textit{credibility} of the detected labels. The credibility is obtained by considering the maximum $p$-value of the given set prediction. Credibility quantifies the quality of the new data points. 

\begin{table}[t]
\centering
\caption{Adoption of confidence for risk-aware ranking on Trust-Hub chip-level Trojan dataset.}
\begin{tabular}{llll}
\hline
  & \textbf{confidence} & \textbf{credibility} & \textbf{y\_pred} \\ \hline
1 & 0.997               & 0.319                & TF               \\ \hline
2 & 0.994               & 0.242                & TF               \\ \hline
3 & 0.922               & 0.162                & TF               \\ \hline
4 & 0.886               & 0.119                & T-EV             \\ \hline
5 & 1                   & 0.645                & TF               \\ \hline
6 & 0.999               & 0.97                 & T-EV             \\ \hline
7 & 0.998               & 0.301                & TF               \\ \hline
\end{tabular}
\label{tab:confcred}
\end{table}


\subsection{Calibrated Explanations for Reject}
\label{Sec:explain}
When the model is not able to detect evolving HT, the model simply says, \textit{``I don't know''} by giving a \textit{NULL} set as the output. In a risk-sensitive domain, a model with no output is better than a decision that is not confident. Our framework also provides the reason for rejecting the decision with a calibrated explanation, as shown in Fig. \ref{fig:lime}, which is different from traditional explainable methods. If we apply a significance level of 0.5 to the given circuit, none of the $p$-values for TF (0.45), TI (0.32), and T-EV (0.23) exceed the significance level. As a result, we reject the decision. The explanation for this rejection is based on Local Interpretable Model-agnostic Explanations (LIME) \cite{dieber2020model}. However, the differentiating factor as compared to SHAP (which disregards causality and is affected by human bias) is that before providing any explanation for the rejection, we ensure that it is calibrated. The approach begins by creating modified versions of the original instance called perturbed instances, where small random changes are introduced. Conformal prediction is then utilized to create prediction regions that estimate the reliability or confidence level of the explanations, and LIME is then used again on these perturbed instances to generate explanations for each of them. The prediction regions obtained through conformal prediction act as a calibration mechanism, guaranteeing that the explanations accurately reflect their level of reliability. 

\section{Conclusion}
\label{Sec:Conclusion}
In this paper, we addressed one of the most neglected evaluation metrics to quantify the predictions made by ML methods in the context of detecting hardware Trojans. First, we designed a method to generate a quality evolving dataset using conformalized generative adversarial network. Then, we proposed an algorithm-agnostic framework called \textbf{PALETTE} to detect evolving hardware Trojans with guaranteed coverage. We also implemented a novel method for rejecting a decision by proving a calibrated explanation. \textbf{PALETTE} is efficient in detecting hardware Trojans with an assigned uncertainty quantification for each detection. 

Our results highlighted opportunities for researchers in related hardware security domains such as logic locking \cite{Rezaei:BreakUnroll, Rezaei:PUF, Maynard:DK-Lock, Aghamohammadi:CoLA} to rethink the application of ML-based solutions and re-construct the metrics to evaluate their methods. We do believe that there is no silver bullet for a zero-day attack, but a robust method to minimize the chances of an attack and a proactive approach to defending the attack do help.

\section*{Acknowledgment}
This work is supported by the National Science Foundation under Award No. 2245247.

\bibliographystyle{IEEEtran}
\bibliography{IEEE}

\end{document}

